\documentclass{ifacconf}

\usepackage{graphicx}      
\usepackage{natbib}        
\usepackage{amsmath}
\usepackage{amsfonts}
\usepackage[dvipsnames]{xcolor}
\begin{document}
\begin{frontmatter}

\title{Exponentially Stable Adaptive Control of MIMO Systems with Unknown Control Matrix} 

\thanks[footnoteinfo]{This research was in part financially supported by Grants Council of the President of the Russian Federation (project MD-1787.2022.4).\\
© 2023 the authors. This work has been accepted to IFAC for publication under a Creative Commons Licence CC-BY-NC-ND}

\author[First]{A. Glushchenko} 
\author[Second]{K. Lastochkin} 

\address[First]{V.A. Trapeznikov Institute of Control Sciences of Russian Academy of Sciences, Moscow, Russia (aiglush@ipu.ru)}
\address[Second]{V.A. Trapeznikov Institute of Control Sciences of Russian Academy of Sciences, Moscow, Russia (lastconst@ipu.ru)}

\begin{abstract}                
The scope of this research is a problem of direct model reference adaptive control of linear time-invariant multi-input multi-output (MIMO) plants without any \textit{a priori} knowledge about system matrices. To handle it, a new method is proposed, which includes three main stages. Firstly, using the well-known DREM procedure, the plant parametrization is made to obtain the linear regressions, in which the plant matrices and state initial conditions are the unknown parameters. Secondly, such regressions are substituted into the known equations for the controller parameters calculation. Thirdly, the controller parameters are identified using the novel \textcolor{black}{adaptive} law with the exponential rate of convergence. To the best of the authors’ knowledge, such a method is the first one to provide the following features simultaneously: 1) it is applicable for the unknown MIMO systems (e.g. without any information about state or control allocation matrices, the sign of the latter, etc.); 2) it guarantees the exponential convergence of both the parameter and tracking errors under the mild requirement of the regressor finite excitation; 3) it ensures element-wise monotonicity of the transient curves of the control law parameters matrices. The results of the conducted experiments with the model of a rubber and ailerons control of a small passenger aircraft corroborate all theoretical results.
\end{abstract}

\begin{keyword}
exponential stability, MIMO systems, unknown control matrix, regressor extension and mixing, finite excitation.
\end{keyword}

\end{frontmatter}

\section{Introduction}
The theory of Model Reference Adaptive Control (MRAC) was developed in the late 1950s as an approach for efficient and safe aircraft control under the condition of significant parameter uncertainty  \citep{c1,c2,c3}. Despite the difficulties, which were faced at the initial stage \citep{c2}, the basic MRAC principles were approved by the engineers over time and found real practical applications in aircraft engineering as well as in other branches of industry \citep{c4}. However, many of the conventional assumptions of the classical MRAC paradigm are still not fully relaxed to this day. This keeps the scientific community interested in the improvement of some aspects of this theory \citep{c3, c5}.

In this research, we consider the requirement to know some {\it a priori} information about the control input (allocation) matrix $B$ for the implementation of the conventional adaptive laws of the form $\Gamma e_{ref}^{\rm{T}}Pf\left( B \right)$.

It is well known \citep{c1, c5} that, in order to be implementable, such adaptive laws require to know either the control allocation matrix $f\left( B \right) = B$, its sign $f\left( B \right) = {\rm{sgn}}\left( B \right)$ or $f\left( B \right) = B_{0}sign(\Lambda)$ for a matrix case. It is rather restrictive for a number of practical scenarios. Therefore, there have been some attempts \citep{c6, c7, c8, c9, c10, c11} to minimize such {\it a priori} needed data about matrix $B$ \textcolor{black}{for MRAC schemes}.

In \citep{c1, c6} conditions are stated, under which the substitution $f\left( B \right) = {B_{ref}}$ is allowed, where ${B_{ref}}$ is the input matrix of the reference model. The method in \citep{c6} leads to a modified control law, which provides only local stability, and requires knowledge of the lower bound of the determinant of the feedforward controller parameter matrix due to the need for its inversion. The identification-based approach $f\left( B \right) = \hat B$ , which is proposed by \citep{c7}, does not require any {\it a priori} information about the matrix $B$, but needs the plant state matrix $A$ to be known instead. This is an even more restrictive assumption for most applications. \citep{c8}, inspired by \citep{c7}, have proposed a combined adaptive control scheme, in which the minimum singular value of the matrix $B$ is assumed to be known. The advantages of \citep{c7, c8} in comparison with the previous solutions \citep{c1, c6} are the relaxation of the regressor persistent excitation (PE) requirement, which is used in \citep{c1, c6}, for the exponential stability of the control scheme and parameter convergence. The method in \citep{c8} includes a nonlinear operator, which prevents singularity when the controller parameters matrices are inverted. However, the finiteness of the number of switches in the course of the adaptation has not been rigorously proved for such an operator, and as a consequence, the chattering is possible.

Considering the adaptive output feedback control problem, the method, which is based on the dynamic regressor extension and mixing (DREM) procedure \citep{c12}, is proposed in \citep{c9}. It guarantees no more than one switch of the nonlinear operator under the condition of a known lower bound of the high-frequency gain. Later this approach has been improved and applied to the problem of the state feedback control of the multi-input multi-output (MIMO) systems in \citep{c10}, in which it is also guaranteed that no more than one switch is required. Such a method needs the lower bound value of the determinant of the control allocation matrix. In \citep{c11}, considering single-input single-output (SISO) systems, another direct adaptive law is proposed, which is based on the I-DREM procedure \citep{c13}. It does not require any {\it a priori} information about $B$ and guarantees the exponential stability of the closed-loop system if the regressor is finitely exciting (FE).

The present paper is an attempt to extend the result of \citep{c11} to the case of MIMO systems. 
The main scientific contribution of the research is threefold: 1) an adaptive control scheme for unknown MIMO plants without any {\it a priori} knowledge about system matrices is obtained; 2) the exponential stability of the closed-loop control system is guaranteed when the vector of signals (states and controls) measured from the plant is finitely exciting (FE); 3) the monotonicity of transients of all control law matrices elements is provided. To the best of the authors’ knowledge, properties 1-3 are provided simultaneously for the first time.

The paper is organized as follows. Section II presents a problem statement. Section III contains the main result. Section IV is to analyze the control system stability. The numerical experiments are presented in Section V.

\textbf{\emph{Notation.}} $\lvert{.}\rvert$ is the absolute value, $\lVert{.}\rVert$ is the Euclidean norm of a vector, $\lambda_{\min }\left( . \right)$ and $\lambda_{\max }\left( . \right)$ are the matrix minimum and maximum eigenvalues, respectively, $vec\left( . \right)$ stands for the matrix vectorization. $I_{n \times n}$, $0_{n \times n}$ and $1_{n \times n}$ are identity, zero and ones $n \times n$ matrices, respectively. We also use the fact that for all (possibly singular) ${n \times n}$ matrices $M$ the following holds: $adj \{M\} M = det \{M\}I_{n \times n}$. The following definition from \textcolor{black}{\citep{c1}} is also used.

\textbf{\emph{Definition 1.}}\emph{ The regressor $\varphi\left(t \right)$ is finitely exciting 
$\left({\varphi\left(t \right)\in{\rm{FE}}}\right)$ over $\left[{t_r^+; \; {\rm{ }}{t_e}}\right]$ if there exist ${t_e} > t_r^ + \ge 0$  and $\alpha  > 0$ such that the following holds 
\begin{equation}\label{eq1}
  \begin{gathered} 
\int\limits_{t_r^ + }^{{t_e}} {\varphi \left( \tau  \right){\varphi ^{\rm{T}}}\left( \tau  \right)d} \tau  \ge \alpha {I_{n \times n}},
  \end{gathered}
\end{equation}
where $\alpha$  is the excitation level.}

Let the corollary of Theorem 9.4 from \textcolor{black}{\citep{c1}} be introduced.

\textbf{\emph{Corollary 1.}}\emph{ For any matrix $D > 0$, a Hurwitz matrix $A \in {\mathbb{R}^{n \times n}}$, a matrix $B$ such that a pair $(A, B)$ is controllable there exists a matrix $P = P^{\rm{T}} > 0$, a scalar $\mu > 0$ and matrices $Q{\rm{,}}\; K $ of appropriate dimension such that:}
\begin{equation}\label{eq2}
  \begin{gathered}
\begin{array}{c}
{A^{\rm{T}}}P + PA =  - Q{Q^{\rm{T}}} - \mu P{\rm{, }} \; PB = QK{\rm{, }}\\
{K^{\rm{T}}}K = D + {D^{\rm{T}}}.
\end{array}
  \end{gathered}
\end{equation}

\section{Problem Statement}
Let the problem of adaptive state feedback control of the linear time-invariant (LTI) MIMO systems be considered:
\begin{equation}\label{eq3}
    \begin{array}{c}
\dot x\left( t \right) = \theta _{AB}^{\rm{T}}\Phi \left( t \right) = Ax\left( t \right) + Bu\left( t \right){\rm{, }} \; x\left( 0 \right) = {x_0}{\rm{,}}\\
\Phi \left( t \right) = {\begin{bmatrix}
{{x^{\rm{T}}}\left( t \right)}&{{u^{\rm{T}}}\left( t \right)}
\end{bmatrix}^{\rm{T}}}{\rm{, }}\;
\theta _{AB}^{\rm{T}} = \begin{bmatrix}
A&B
\end{bmatrix}
{\rm{,}}
\end{array}
\end{equation}
where $x\left( t \right) \in {\mathbb{R}^n}$ is the measurable state vector, \linebreak ${x_0} \in {\mathbb{R}^n}$ is the \textcolor{black}{known (or possibly unknown)} vector of initial conditions, ${u\left( t \right)} \in {\mathbb{R}^m}$ is a control vector, ${A} \in {\mathbb{R}^{n \times n}}$ is the state matrix, ${B} \in {\mathbb{R}^{n \times m}}$ is the control (allocation) matrix of full column rank. The pair $\begin{pmatrix}A,\; B\end{pmatrix}$
is controllable. The vector ${\Phi\left( t \right)} \in {\mathbb{R}^{n + m}}$ is considered to be measurable at each time instant $t>0$, whereas ${\theta _{AB}} \in {\mathbb{R}^{\left( {n + m} \right) \times n}}$ is time-invariant and unknown.

The reference model, which is used to define the required control quality for \eqref{eq3}, is described as:
\begin{equation}\label{eq4}
{\dot x_{ref}}\left( t \right) = {A_{ref}}{x_{ref}}\left( t \right) + {B_{ref}}r\left( t \right){\rm{, }}\;{x_{ref}}\left( 0 \right) = {x_{0ref}}{\rm{,}}
\end{equation}
where ${x_{ref}\left( t \right)} \in {\mathbb{R}^{n}}$ is the reference model state vector, ${x_{0ref}}$ is the initial conditions vector, $r\left( t \right) \in {\mathbb{R}^m}$ is the reference, ${A _{ref}} \in {\mathbb{R}^{n \times n}}$ is the Hurwitz state matrix of the reference model, ${B _{ref}} \in {\mathbb{R}^{n \times m}}$ is the input matrix.

The matching condition is supposed to be met for the plant \eqref{eq3} and the reference model \eqref{eq4}.

\textbf{\emph{Assumption 1.}}\emph{ There exist ${K_x} \in {\mathbb{R}^{m \times n}}$ and ${K_r} \in {\mathbb{R}^{m \times m}}$ such that the following holds:}
\begin{equation}\label{eq5}
{A + B{K_x} = {A_{ref}}{\rm{, }}\;\;B{K_r} = {B_{ref}}.}
\end{equation}

Considering \eqref{eq5}, the control law for \eqref{eq3} is chosen as:
\begin{equation}\label{eq6}
{u\left( t \right) = {\hat K_x}\left( t \right)x\left( t \right) + {\hat K_r}\left( t \right)r\left( t \right){\rm{,}}}
\end{equation}
where ${\hat K_x}\left( t \right) \in {\mathbb{R}^{m \times n}}$ and ${\hat K_r}\left( t \right) \in {\mathbb{R}^{m \times m}}$ are adjustable parameters and ${\hat K_r}\left( 0 \right) \ne {0_{m \times m}}$.

Having substituted \eqref{eq6} into \eqref{eq3} and subtracted \eqref{eq4} from the obtained equation, the following error equation is written:
\begin{equation}\label{eq7}
{{\dot e_{ref}}\left( t \right) = {A_{ref}}{e_{ref}}\left( t \right) + B\left( {{{\tilde K}_x}\left( t \right)x\left( t \right) + {{\tilde K}_r}\left( t \right)r} \right).}
\end{equation}

Here ${e_{ref}\left( t \right)}\! \buildrel \Delta \over =\! x\left( t \right) \!-\! {x_{ref}\left( t \right)}{\rm{, }} \; {\tilde K_x}\left( t \right)\! \buildrel \Delta \over =\! {\hat K_x}\left( t \right)\! -\! {K_x}{\rm{, }} \; {\tilde K_r}\left( t \right) \buildrel \Delta \over = \linebreak \buildrel \Delta \over =  {\hat K_r}\left( t \right) - {K_r}$. The following notation is introduced in \eqref{eq7}:
\begin{equation}\label{eq8}
{\begin{array}{c}
{{\tilde \theta }^{\rm{T}}}\left( t \right){\rm{ = }}

\begin{bmatrix}
{{{\tilde K}_x}\left( t \right)}&{{{\tilde K}_r}\left( t \right)}
\end{bmatrix}

= {{\hat \theta }^{\rm{T}}}\left( t \right) - {\theta ^{\rm{T}}}{\rm{,}}\\

\omega \left( t \right) = 

\begin{bmatrix}
{{x^{\rm{T}}}\left( t \right)}&{{r^{\rm{T}}}\left( t \right) }
\end{bmatrix}

^{\rm{T}}{\rm{,}}
\end{array}}
\end{equation}
where  $\omega\left( t \right)   \in {\mathbb{R}^{n + m}}$, $\hat \theta  \in {\mathbb{R}^{\left( {n + m} \right) \times m}}$ is the vector of the control law adjustable parameters.

Considering \eqref{eq8} and the difference between the initial conditions for \eqref{eq3} and \eqref{eq4}, the equation \eqref{eq7} is written as:
\begin{equation}\label{eq9}
{{\dot e_{ref}}\left( t \right)\! =\! {A_{ref}}{e_{ref}}\left( t \right)\! + \!B{\tilde \theta ^{\rm{T}}}\left( t \right)\omega \left( t \right){\rm{, }}\;{e_{ref}}\left( 0 \right)\! =\! {e_{0ref}}.}
\end{equation}

The objective is formulated on the basis of \eqref{eq9}.

\textbf{\emph{Goal.}}\emph{ When $\Phi\left( t \right) \in {\rm{FE}}$, it is required to hold:}
\begin{equation}\label{eq10}
{\mathop {{\rm{lim}}}\limits_{t \to \infty } \left\| {\xi \left( t \right)} \right\| = 0{\rm{ }}\left( {exp} \right){\rm{,}}}
\end{equation}
{\it where $\xi\left( t \right)  = {
\begin{bmatrix}
{e_{ref}^{\rm{T}}}\left( t \right) &{ve{c^{\rm{T}}}\left( {\tilde \theta\left( t \right)  } \right)}
\end{bmatrix}^{\rm{T}}} \in {\mathbb{R}^{n + m\left( {n + m} \right)}}$ is the augmented tracking error.}

\section{Main Result}
The main result of this study is based on the {\it direct self-tuning regulators} concept. Having $\Phi\left( t \right) $ at hand, it is proposed to apply some mathematical transformations to $\left(\ref{eq3}\right)$ and, using the obtained results, derive the measurable regression equation with the help of $\left(\ref{eq5}\right)$:
\begin{equation}\label{eq11}
    \begin{array}{c}
{y_\theta }\left( t \right) = \Delta \left( t \right)\theta {\rm{, }}	
\end{array}
\end{equation}
where $\Delta\left( t \right)  \!\in\! {\mathbb{R}},\; y_\theta\!\left( t \right) \!\in\! {\mathbb{R}^{\left( {n + m} \right) \times m}}$ \textcolor{black}{are} measurable signals.

Applying the results of \citep{c11, c13}, Corollary 1 and conditions of the exponential stability, it is proposed to derive the adaptive law for $\hat\theta\left( t \right) $ on the basis of $\left(\ref{eq11}\right)$, which guarantees that $\left(\ref{eq10}\right)$ holds when $\Phi\left( t \right)  \in {\rm{FE}}$.

The subsection $3.1$ of this section contains the description of steps to obtain the regression equation $\left(\ref{eq11}\right)$,  under the condition that the plant $\left(\ref{eq3}\right)$ matrices are unknown and  $\dot x\left( t \right) $ is unmeasurable. The subsection $3.2$ presents the adaptive law, which ensures that $\left(\ref{eq10}\right)$ is achieved and does not require {\it a priori} knowledge of the plant $\left(\ref{eq3}\right)$ matrices.

\subsection{Parameterization}
Let the stable filters of the variables of $\left(\ref{eq3}\right)$ be introduced:
\begin{equation}\label{eq12}
    \begin{array}{c}
\dot{\overline{\mu}}\left( t \right)  =  - l\overline{\mu}\left( t \right) + \dot x\left( t \right){\rm{,}} \; \overline{\mu}\left(0\right) = {0_n}{\rm{, }} \\
\dot{\overline{\Phi}}\left( t \right)   =  - l\overline{\Phi}\left( t \right)  + \Phi\left( t \right) {\rm{, }} \; \overline{\Phi}\left( 0 \right) = {0_{n + m}}{\rm{,}}
\end{array}
\end{equation}
where $l > 0$ is the filter constant.

The regressor $\overline{ \Phi}\left( t \right) $ is calculated as a solution of the second equation of $\left(\ref{eq12}\right)$, whereas, according to \citep{c10}, $\mu\left( t \right) $ is calculated without the value of  $\dot x\left( t \right) $:
\begin{equation}\label{eq13}
    \begin{array}{c}
\overline{\mu} \left( t \right) = {e^{ - lt}}\left( {\overline{\mu} \left( 0 \right) - x\left( 0 \right) + l\overline{x}\left( 0 \right)} \right) + x\left( t \right) - l\overline{x}\left( t \right){\rm{,}}
\end{array}
\end{equation}
where $\overline{x}\left( t \right)\!=\!\begin{bmatrix}
I_{n \times n} &0_{n \times m}
\end{bmatrix}\overline{\Phi}\left( t \right) $ are first $n$ elements of $\overline{\Phi}\left( t \right) $.

Considering $\left(\ref{eq12}\right)$ and unmeasurable initial conditions $x\left( 0 \right)$ in $\left(\ref{eq13}\right)$, the equation $\left(\ref{eq3}\right)$ is rewritten as\footnote{\textcolor{black}{If $x(0)$ is known, then $\overline{\varphi}\left( t \right)$ and $\overline{ \theta} _{AB}$ do not contain $e^{ - lt}$ and $x(0)$, respectively.}}:
\begin{equation}\label{eq14}
\begin{array}{c}
\overline{z}\left( t \right) = \overline{\mu} \left( t \right) + {e^{ - lt}}x\left( 0 \right) = \overline{\theta} _{AB}^{\rm{T}}\overline{\varphi} \left( t \right){\rm{,}}\\
\overline{\varphi}\left( t \right) = 
\begin{bmatrix}
\overline{\Phi }^{\rm{T}}\left( t \right) & {{e^{ - lt}}}
\end{bmatrix}^{\rm{T}}{\rm{, }} \;
\overline{ \theta} _{AB}^{\rm{T}} =
\begin{bmatrix}
A&B&{x\left( 0 \right)}
\end{bmatrix}{\rm{, }} 
\end{array}
\end{equation}
where $\overline{z}\left(t\right)\!=\!x\left(t\right)\,-\,l\overline{x}\left( t \right) $ is a measurable function $\forall t > 0$, $\overline{\varphi}\left( t \right)  { \in {\mathbb{R}}^{n + m + 1}}$ is a measurable regressor, \linebreak $\overline{\theta} _{AB}{ \in {\mathbb{R}}^{\left( {n + m + 1} \right) \times n}}$ is an extended vector of the unknown parameters.

\textbf{\emph{Assumption 2.}}\emph{ The parameter $l$ is chosen so as the implication $\overline{\Phi}\left( t \right)   \in {\rm{FE}}$ $\Rightarrow{\overline{\varphi}}\left( t \right)  \in {\rm{FE}}$ holds.}

Let the minimum-phase operator $\mathfrak{H}\left[ . \right]{\rm{:}} = {1 \mathord{\left/
 {\vphantom {1 {\left( {p + k} \right)}}} \right.
 \kern-\nulldelimiterspace} {\left( {p + k} \right)}}\left[ . \right]$ be introduced $\left( p:= {d \over dt} \right)$. Then, the DREM technique \citep{c12} can be applied to $\left(\ref{eq14}\right)$:
\begin{equation}\label{eq15}
\begin{array}{c}
z\left( t \right) = \varphi \left( t \right){{\overline{\theta}}_{AB}}{\rm{,}}\\
z\left( t \right){\rm{:}}  = {\rm{adj}}\left\{ {\mathfrak{H}\left[ {\overline{\varphi} \left( t \right){{\overline{\varphi}}^{\rm{T}}}\left( t \right)} \right]} \right\} \mathfrak{H}\left[ {\overline{\varphi} \left( t \right){{\overline{z}}^{\rm{T}}}\left( t \right)} \right]{\rm{, }}\\
\varphi \left( t \right){\rm{:}}  = {\rm{det}} \left\{ {\mathfrak{H}\left[ {\overline{\varphi} \left( t \right){{\overline{\varphi} }^{\rm{T}}}\left( t \right)} \right]} \right\}{\rm{,}}
\end{array}
\end{equation}
where $k > 0,\; \varphi\left( t \right)   \in {\mathbb{R}}, \; z\left( t \right) { \in {\mathbb{R}}^{\left( {n + m + 1} \right) \times n}}$.

The following regressions are obtained from $\left(\ref{eq15}\right)$:
\begin{equation}\label{eq16}
\begin{gathered}
{z_A}\left( t \right) = {z^{\rm{T}}}\left( t \right)\mathfrak{L} = \varphi \left( t \right)A{\rm{,  }}\\
{z_B}\left( t \right) = {z^{\rm{T}}}\left( t \right){\mathfrak{e}_{n + m + 1}} = \varphi \left( t \right)B{\rm{,}}
\end{gathered}
\end{equation}
where $\mathfrak{L}=\begin{bmatrix}\begin{smallmatrix} {I_{n \times n}} & {0_{n \times \left( {m + 1} \right)}} \end{smallmatrix}\end{bmatrix}^{\rm{T}}{ \in {\mathbb{R}}^{\left( {n + m + 1} \right) \times n}}$, $\mathfrak{e}_{n + m + 1}=\linebreak =\begin{bmatrix}\begin{smallmatrix}{0_{m \times n}} & {I_{m \times m}} & 0_{m \times 1} \end{smallmatrix}\end{bmatrix}^{\rm{T}}{ \in {\mathbb{R}}^{(n + m + 1)\times m}}$.

The main benefit of DREM application is that the regressions in \eqref{eq16} have scalar regressors, so, having multiplied \eqref{eq5} by $\varphi\left( t \right)$, we can substitute \eqref{eq16} into \eqref{eq5} to obtain:
\begin{equation}\label{eq17}
\begin{gathered}
{{\overline{y}}_\theta }\left( t \right) \!{\rm{:}} =\!
\begin{bmatrix} {\varphi \left( t \right){A_{ref}} - {z_A}\left( t \right)} & \;\;{\varphi \left( t \right){B_{ref}}}\end{bmatrix}^{\rm{T}}\!=\! \theta z_B^{\rm{T}}\left( t \right),
\end{gathered}
\end{equation}
where ${\overline{y}_\theta }\left( t \right) { \in {\mathbb{R}^{\left( {n + m} \right) \times n}}{\rm{, }} \; z_B^{\rm{T}}\left( t \right)  { \in {\mathbb{R}^{m \times n}}}}$.

It should be noted that the above-mentioned substitution is not possible without regression scalarization \eqref{eq15}.

The equation $\left(\ref{eq17}\right)$ is transposed and multiplied by ${\rm{adj}}\left\{ {z_B^{\rm{T}}\left( t \right) z_B\left( t \right)} \right\}z_B^{\rm{T}}\left( t \right) $ to obtain $\left(\ref{eq11}\right)$ exact to notation:

$$\begin{array}{c}
\Delta \left( t \right){\rm{:}} = {\rm{det}} \left\{ {z_B^{\rm{T}}\left( t \right) z_B\left( t \right)}  \right\},\\ 
y_\theta ^{\rm{T}}\left( t \right){\rm{:}} = {\rm{adj}}\left\{ {z_B^{\rm{T}}\left( t \right) z_B\left( t \right)} \right\}z_B^{\rm{T}}\left( t \right)\overline{ y}_\theta ^{\rm{T}}\left( t \right).
\end{array}$$

\textbf{\emph{Remark 1.}}\emph{ As the filters \eqref{eq12} are stable, if  $\Phi\left( t \right)   \in {\rm{FE}}$, then $\overline{\Phi}\left( t \right)   \in {\rm{FE}}$. So, following Assumption 2, these results in $\overline{ \varphi}\left( t \right)   \in {\rm{FE}}$ . In \citep{c14} the implication \linebreak $\overline{ \varphi}\left( t \right)   \in {\rm{FE}} \Rightarrow \varphi\left( t \right)   \in {\rm{FE}}$  is proved for DREM \eqref{eq15}. Then, as the regressor $\Delta\left( t \right) $ depends on only one variable $\varphi\left( t \right) $, and the matrix $B$ has full column rank $\left( {\det \left( {{B^{\rm{T}}}B} \right) > 0} \right)$, it also holds that  $\varphi\left( t \right)   \in {\rm{FE}} \Rightarrow \Delta\left( t \right)   \in {\rm{FE}}$}.
\subsection{Adaptive Law}
The adaptive law of the control law parameters \eqref{eq6} is introduced on the basis of the regression \eqref{eq11}:
\begin{equation}\label{eq18}
\dot{\hat{ \theta}} \left( t \right) =  - \gamma \Delta \left( t \right)\left( {{y_\theta }\left( t \right) - \Delta \left( t \right)\hat \theta } \right){\rm{,}}
\end{equation}
where  $\gamma  > 0$ is the adaptive gain.

However, only instantaneous data (see \citep{c15}) are used in \eqref{eq18}. So, based on proof from \citep{c12}, the law \eqref{eq18} provides exponential convergence of $\tilde{\theta}\left( t \right) $ to zero only when $\Delta\left(t\right)\in~{\rm{PE}}$. This does not satisfy the objective \eqref{eq10}. To this end, to achieve \eqref{eq10}, the adaptive law is to be derived on the basis of one of the known approaches \citep{c13, c15, c16, c17}, which relax $\Delta\left( t \right)   \in {\rm{PE}}$ to $\Delta\left( t \right)  \in {\rm{FE}}$.

In this study, it is proposed to choose the one described in \citep{c13}. According to it, let the exponential filter with forgetting $\mathfrak{G}\left[ . \right]{\rm{:}} = {{{e^{ - \sigma t}}} \mathord{\left/
 {\vphantom {{{e^{ - \sigma t}}} p}} \right.
 \kern-\nulldelimiterspace} p}\left[ . \right]$ be introduced and applied to the regression \eqref{eq11} to obtain:
 \begin{equation}\label{eq19}
 {\Upsilon _\theta }\left( t \right){\rm{:}} = \mathfrak{G}\left[ {\Delta \left( t \right){y_\theta }\left( t \right)} \right] = \mathfrak{G}\left[ {{\Delta ^2}\left( t \right)} \right]\theta  = \Omega \left( t \right)\theta {\rm{,}}
 \end{equation}
where $\sigma  > 0$ is the parameter of the operator $\mathfrak{G}\left[ . \right]$.

The following holds for the new regressor $\Omega\left( t \right) $.

\textbf{\emph{Proposition 1.}}\emph{ If $\Delta\left( t \right)   \in {\rm{FE}}$  over $\left[ {t_r^+{\rm{;}}\; {t_e}} \right]$ and $\left( \Delta\left( t \right)   \in {L_\infty }\right.$ or  $\left. {\left| \Delta\left( t \right)   \right| \le {c_1}{e^{{c_2}t}}\;and\;\sigma  > {\rm{2}}{c_2}{\rm{,}}\;{c_1} > 0,\;{c_2} > 0} \right)$, then:
\begin{enumerate}
\item$\forall t \ge t_r^+ \;\;\; \Omega \left( t \right) \in {L_\infty }{\rm{,} \;\; }\Omega \left( t \right) \ge 0;$
\item $\forall t \ge {t_e} \;\;\; \Omega \left( t \right) > 0,\;\;{\Omega _{LB}} \le \Omega \left( t \right) \le {\Omega _{UB}}.$
\end{enumerate}}

{\it Proof of Proposition 1 is presented in \citep{c11,c17i5} (Section 2).}

Using Proposition 1, the adaptive law to guarantee the exponential convergence of  $\tilde \theta\left( t \right)$ when $\Phi\left( t \right)  \in {\rm{FE}}$, is:
 \begin{equation}\label{eq20}
 \dot{\hat{\theta}} \left( t \right) =  - \gamma \Omega \left( t \right)\left( {\Omega \left( t \right)\hat \theta  - {\Upsilon _\theta }\left( t \right)} \right){\rm{ = }} - \gamma {\Omega ^2}\left( t \right)\tilde \theta \left( t \right).
 \end{equation}
 
\section{Stability Analysis and Some Remarks}
According to \citep{c13}, the adaptive law \eqref{eq20} provides exponential convergence of the parameter error $\tilde \theta\left( t \right)$ only. In the following theorem we formulate strict formal conditions, under which the law \eqref{eq20} guarantees that the objective \eqref{eq10} is met.

\textbf{\emph{Theorem 1.}}\emph{ Let $\Phi\left( t \right)  \in {\rm{FE}}$, then, if $\gamma$ value is chosen according to}
\begin{equation}\label{eq21}
{\gamma  \buildrel \Delta \over = \left\{ \begin{array}{l}
{\rm{0}}{\rm{,\;\; if\;\;}}\Omega \left( t \right) = 0,\\
{\textstyle{{{\gamma _0}{\lambda _{{\rm{max}}}}\left( {\omega \left( t \right){\omega ^{\rm{T}}}\left( t \right)} \right) + {\gamma _1}} \over {{\Omega ^2}\left( t \right)}}}{\rm{\;\;otherwise}}{\rm{,}}
\end{array} \right.}
\end{equation}
{\it then the adaptive law \eqref{eq20} ensures that:}
\begin{enumerate}
\item $\forall {t_a} \ge {t_b}\;\;{\rm{ }}\left| {{{\tilde \theta }_i}\left( {{t_a}} \right)} \right| \le \left| {{{\tilde \theta }_i}\left( {{t_b}} \right)} \right|{\rm{;}}$
\item $\forall t \ge t_r^ + \;\; {\rm{ }}\xi\left( t \right)  \in {L_\infty }{\rm{;}}$
\item $\forall t \ge {t_e}$ {\it the error $\xi$ converges exponentially to zero at the rate, which minimum value is directly proportional to the parameters ${\gamma _0} \ge 1$ and ${\gamma _1} \ge 0$.}
\end{enumerate}

{\it Proof of Theorem is presented in Appendix.}

Thus, in contrast to existing approaches \citep{c5,c6,c7,c8,c9,c10,c11}, when $\gamma$ is chosen according to \eqref{eq21}, the adaptive law \eqref{eq20} does not require any {\it a priori} information about the matrices of the system \eqref{eq3} and guarantees that the objective \eqref{eq10} is fulfilled. Some additional technical details of the law \eqref{eq20} implementation can be found in \citep{c11}.

\textbf{\emph{Remark 2.}}\emph{ When $\Phi\left( t \right)  \in {\rm{FE}}$, the switching of the nonlinear operator \eqref{eq21} may occur only once, since, according to Proposition 1, $\Omega\left( t \right)  $ is a positive semidefinite function.}

\textbf{\emph{Remark 3.}}\emph{ The condition $\Phi \left( t \right)  \in {\rm{FE}}$ is necessary but not sufficient to achieve \eqref{eq10}. According to Assumption 2 and Remark 1, the necessary and sufficient conditions are $\overline{\varphi}\left( t \right)   \in {\rm{FE}}$ and Assumption 2. This may become a dramatically critical requirement for some applications.}

\textbf{\emph{Remark 4.}}\emph{ In contrast to the baseline direct laws of the form $\Gamma e_{ref}^{\rm{T}}\left( t \right) Pf\left( B \right)$, the proposed one \eqref{eq20} provides stability of ${e_{ref}}\left( t \right) $  only when $\Phi\left( t \right)   \in {\rm{FE}}$. Therefore, to implement \eqref{eq20} in practice, the a priori information is required that this condition holds. If $\Phi\left( t \right)  \notin {\rm{FE}}$ for a particular plant \eqref{eq3} and a particular reference, then it is possible to hold $\Phi\left( t \right)   \in {\rm{FE}}$ artificially by addition of the dither noise to the control or the reference signal according to \citep{c18,c19}. At the same time, we can not use conventional modular design \citep{c21} to ensure stability of the control system when $t<t_e$ as neither the sign nor the values of elements of matrix B are known.The actual problem is to obtain conditions on $r\left( t \right)$, under which the requirement $\Phi\left( t \right)   \in {\rm{FE}}$ is met for the whole class of MIMO plants \eqref{eq3}.}

\section{Numerical Simulations}
Numerical simulation to test the proposed adaptive control system, which consists of the control law \eqref{eq6}, processing procedure \eqref{eq12}, \eqref{eq13}, \eqref{eq15}-\eqref{eq17}, \eqref{eq19}, and the adaptive law \eqref{eq20}, was conducted using the model of a lateral-directional motion of a conventional small passenger aircraft from \citep{c20}:
\begin{equation}\label{eq22}
\begin{gathered}
\dot x\! =\! \bigg(\begin{smallmatrix}
  0 & 0 & 1 & 0\\
  0.049 & -0.083 & 0 & -1\\
  0 & -4.55 & -1.70 & 0.172\\
  0 & 3.382 & -0.065 & -0.089
\end{smallmatrix}\bigg)\!x \!+\! 
\bigg(\begin{smallmatrix}
  0 & 0\\
  0 & 0.012\\
  27.276 & 0.576\\
  0.395 & -1.362
\end{smallmatrix}\bigg)u, \\
x_0=\small(\begin{matrix}
-1&-0.5&0&0
\end{matrix}\small)^{\rm{T}}.
\end{gathered}
\end{equation}
where $x_1$ is the bank angle, $x_2$ is the sideslip angle, $x_3$ is the roll rate, $x_4$ is the vehicle yaw rate, $u_1$ is the aileron position, $u_2$  is the rudder position. According to the problem statement, all plant \eqref{eq22} parameters and initial conditions were considered as unknown.

The reference model and reference for \eqref{eq22} were also chosen as in \citep{c20}:
\begin{displaymath}
\begin{gathered}
\dot x_{ref} \! =\!\! \bigg(\begin{smallmatrix}
  0 & 0 & 1 & 0\\
  0.048 & -0.082 & 0 & -0.976\\
  -19.53 & -5.219 & -10.849 & 1.822\\
  -0.204 & 3.22 & -0.145 & -2.961
\end{smallmatrix}\bigg)\!x_{ref}\! +\! \!
\bigg(\begin{smallmatrix}
  0 & 0\\
  0 & 0.029\\
  19.441 & 5.317\\
  0.348 & -3.379
\end{smallmatrix}\bigg)\!r, \\
\small
{r_1} = 1,{\rm{\;\;}}{r_2} =\begin{matrix} 0.5\left( {1 - {e^{ - 10t}}} \right)\end{matrix}.
\end{gathered}
 \end{displaymath}

The parameters of the adaptive law \eqref{eq20}, filters \eqref{eq12}, operators $\mathfrak{H}\left[ . \right]{\rm{, }}\mathfrak{G}\left[ . \right]$, and the initial values of the parameters of the control law \eqref{eq6} were picked as follows:
\begin{equation}\label{eq23}
\begin{gathered}
l = 1,{\rm{\;\;}}k = 10,{\rm{\;\;}}{\gamma _1} = 10,{\rm{\;\;}}{\gamma _0} = 1, \;\; \sigma  = {\textstyle{1 \over 2}},\\ {\hat \theta ^{\rm{T}}}\left( 0 \right) = 
{\begin{bmatrix}
{{0_{m \times n}}}&{{I_{m \times m}}}
\end{bmatrix}}.
\end{gathered}
\end{equation}

As for practical implications of Assumptions 1 and 2, the plant and reference model had the same structure, all plant equations with parametric uncertainty contained sufficient number of controls, \eqref{eq12} ensured excitation propagation.

Figure 1 shows transient curves of the plant \eqref{eq22} states and ideal ones, which were obtained by setting into the reference model the following plant initial conditions: ${x_{ref}}\left( 0 \right) = x\left( 0 \right)$. Figure 2 is to compare the ideal control vector $u^*$ and the one $u$ obtained from the proposed adaptive system. Figure 3 demonstrates transients of the controller feedback parameters ${\hat K_x\left( t \right) }$, whereas Figure 4 – of the feedforward ones ${\hat K_r}\left( t \right) $. In the figures controller parameters estimates were "frozen" till $0.15$ s as till that moment $\Delta(t) = 0$ because condition \eqref{eq1} had not been satisfied yet ($t_{e}=0.15$ s).

   \begin{figure}[thpb]
      \centering
      \includegraphics[scale=0.75]{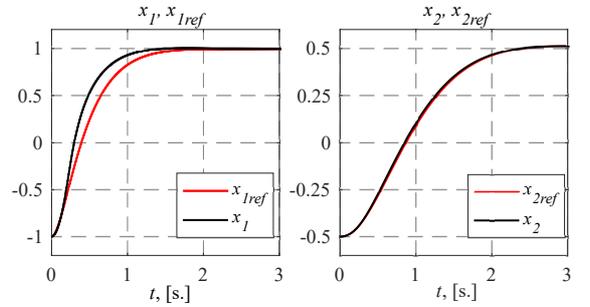}
      \caption{ Behavior of the plant and reference model states.}
      \label{Figure1} 
      \end{figure}
      
      \begin{figure}[thpb]
      \centering
        \includegraphics[scale=0.75]{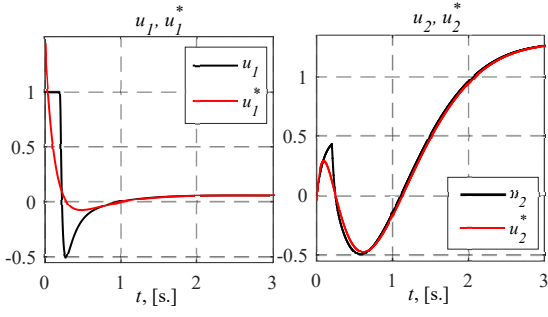}
      \caption{ Ideal $u^*$ and obtained $u$ control signals.}
      \label{Figure2} 
      \end{figure}
      
      \begin{figure}[thpb]
      \centering
      \includegraphics[scale=0.75]{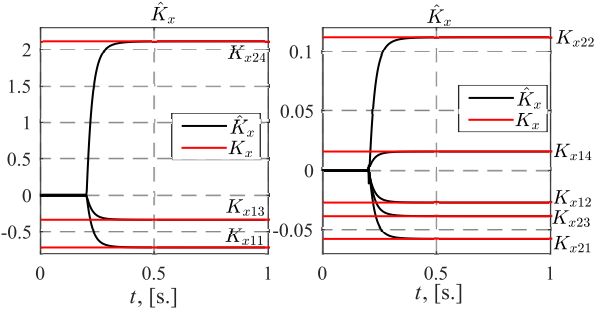}
      \caption{ \textcolor{black}{Behavior} of feedback parameters ${\hat K_x}\left( t \right) $.}
      \label{Figure3}
   \end{figure}

\begin{figure}[thpb]
      \centering
       \includegraphics[scale=0.75]{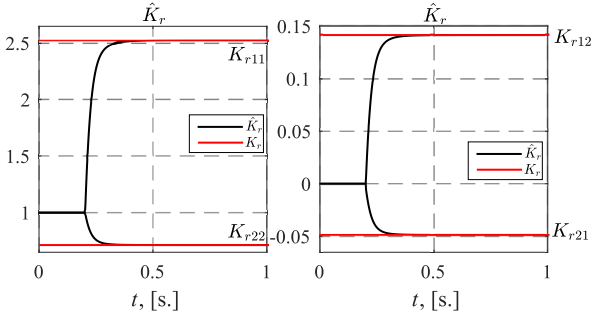}
      \caption{ Transient curves of feedforward parameters ${\hat K_r}\left( t \right) $.}
      \label{Figure4}
   \end{figure}

The results of the numerical experiments presented in Figures 1-4 validated the contribution of this work noted in the introduction in comparison with known solutions \citep{c5,c6,c7,c8,c9,c10,c11}. They confirmed the theoretical conclusions of Theorem 1, and showed that \eqref{eq10} was achieved.

Important features of the developed system are the monotonicity of each controller parameter adjustment process, and the fact that the rate of the parameter convergence is directly adjustable by choice of scalar parameters $\gamma _0$ and $\gamma _1$. These improve the transient behavior of states and control vector, and make the commissioning of the developed adaptive system significantly easier.

\section{Conclusion}
Considering the unknown MIMO plants, the adaptive state feedback control system has been proposed, which does not require any {\it a priori } information about the plant matrices. When the regressor vector $\Phi\left( t \right)$, which is constituted of the plant states and control signals, is finitely exciting, the proposed system guarantees exponential stability of the tracking error, and the exponential convergence of the control law parameters identification error to zero. Considering numerical experiments, the effectiveness of the proposed approach to solve the adaptive control problem of the lateral-directional motion of the conventional small passenger aircraft was demonstrated.

Establishing the conditions, under which $\Phi\left( t \right)$  is FE for any generic MIMO system, is the scope of further research, as well as the plans to extend the obtained results to output-feedback control case.

\renewcommand{\theequation}{A\arabic{equation}}
\setcounter{equation}{0}  

\section*{Appendix}
{\it Proof of Theorem 1.} As $\Omega\left( t \right)   \in \mathbb{R}$, the solution of \eqref{eq20} is:
\begin{equation}\label{eqA1}
{\tilde \theta _i}\left( t \right) = {e^{ - \int\limits_{t_r^+ }^t {\gamma \left( \tau  \right){\Omega ^2}\left( \tau  \right)d\tau } }}{\tilde \theta _i}\left( {t_r^ + } \right).
\end{equation}

As ${\rm{sign}}\left( {\gamma {\Omega ^2\left( t \right) }} \right) = {\mathop{\rm const}\nolimits}  > 0$, then it follows from \eqref{eqA1} that $\left| {{{\tilde \theta }_i}\left( {{t_a}} \right)} \right| \le \left| {{{\tilde \theta }_i}\left( {{t_b}} \right)} \right|{\rm{ }}\forall {t_a} \ge {t_b}$. This completes the proof of the first part of the theorem.

As ${A_{ref}}$ is a Hurwitz matrix and $B$ is of full column rank, then according to the corollary of KYP lemma \eqref{eq2}, there exist $P{\rm{, }}\;Q{\rm{, }}\;D{\rm{, }}\;K,$ which satisfy \eqref{eq3}. Then, the quadratic function to analyze the stability of \eqref{eq9} is chosen:
\begin{equation}\label{eqA2}
\begin{gathered}
V \!=\! e_{ref}^{\rm{T}}P{e_{ref}} \!+\! {\textstyle{1 \over 2}}tr\left( {{{\tilde \theta }^{\rm{T}}}\tilde \theta } \right){\rm{, }}\;H \!=\! {\rm{blockdiag}}\left\{ {P{\rm{,\;}}{\textstyle{1 \over 2}}I} \right\}\\
\underbrace {{\lambda _{{\rm{min}}}}\left( H \right)}_{{\lambda _{\mathop{\rm m}\nolimits} }}{\left\| \xi  \right\|^2} \le V\left( {\left\| \xi  \right\|} \right) \le \underbrace {{\lambda _{{\rm{max}}}}\left( H \right)}_{{\lambda _M}}{\left\| \xi  \right\|^2}{\rm{,}}
\end{gathered}
\end{equation}

Applying $tr\left( {AB} \right) = BA$, the derivative of \eqref{eqA2} with respect to \eqref{eq9}, \eqref{eq20} is written as:
\begin{equation}\label{eqA3}
\begin{array}{c}
\dot V = e_{ref}^{\rm{T}}\left( {A_{ref}^{\rm{T}}P + P{A_{ref}}} \right){e_{ref}} + 2e_{ref}^{\rm{T}}PB{{\tilde \theta }^{\rm{T}}}\omega  - \\
 - tr\left( {{{\tilde \theta }^{\rm{T}}}\gamma {\Omega ^2}\tilde \theta } \right) =  - \mu e_{ref}^{\rm{T}}P{e_{ref}} - \\
 - e_{ref}^{\rm{T}}Q{Q^{\rm{T}}}{e_{ref}} + tr\left( {2{{\tilde \theta }^{\rm{T}}}\omega e_{ref}^{\rm{T}}QK - {{\tilde \theta }^{\rm{T}}}\gamma {\Omega ^2}\tilde \theta } \right).
\end{array}
\end{equation}

Without the loss of generality, $D$ is chosen so as $K{K^{\rm{T}}}~=~{K^{\rm{T}}}K = {I_{m \times m}}$. Then we have:
\begin{equation}\label{eqA4}
\begin{array}{c}
\dot V =  - \mu e_{ref}^{\rm{T}}P{e_{ref}} + tr\left( { - {K^{\rm{T}}}{Q^{\rm{T}}}{e_{ref}}e_{ref}^{\rm{T}}QK + } \right.\\
 + 2{{\tilde \theta }^{\rm{T}}}\omega e_{ref}^{\rm{T}}QK\left. { \pm {{\tilde \theta }^{\rm{T}}}\omega {\omega ^{\rm{T}}}\tilde \theta  - {{\tilde \theta }^{\rm{T}}}\gamma {\Omega ^2}\tilde \theta } \right) = \\
 =  - \mu e_{ref}^{\rm{T}}P{e_{ref}} + tr\left( { - \left( {e_{ref}^{\rm{T}}QK - {{\tilde \theta }^{\rm{T}}}\omega } \right) \times } \right.\\
 \times {\left( {e_{ref}^{\rm{T}}QK - {{\tilde \theta }^{\rm{T}}}\omega } \right)^{\rm{T}}} + \left. {{{\tilde \theta }^{\rm{T}}}\omega {\omega ^{\rm{T}}}\tilde \theta  - {{\tilde \theta }^{\rm{T}}}\gamma {\Omega ^2}\tilde \theta } \right) \le \\
 \le  - \mu e_{ref}^{\rm{T}}P{e_{ref}} + tr\left\{ {{{\tilde \theta }^{\rm{T}}}\omega {\omega ^{\rm{T}}}\tilde \theta  - {{\tilde \theta }^{\rm{T}}}\gamma {\Omega ^2}\tilde \theta } \right\}.
\end{array}
\normalsize
\end{equation}

Two cases are considered: $t < {t_e}$ and $t \ge {t_e}$. Firstly, let $t < {t_e}$. Following \eqref{eq20} and \eqref{eq21}, it holds that $\Omega\left( t \right)   = 0$ and $\left\| {\tilde \theta } \right\| = \left\| {\tilde \theta \left( 0 \right)} \right\|$. Then, $\forall t < {t_e}$ \eqref{eqA4} is rewritten as:
\begin{equation}\label{eqA5}
\begin{array}{c}
\dot V \!\le\!  - \mu e_{ref}^{\rm{T}}P{e_{ref}} + tr\left\{ {{{\tilde \theta }^{\rm{T}}}\left( 0 \right)\omega {\omega ^{\rm{T}}}\tilde \theta \left( 0 \right) \pm {{\tilde \theta }^{\rm{T}}}\tilde \theta } \right\} \le \\
 \le  - \mu e_{ref}^{\rm{T}}P{e_{ref}} - tr\left\{ {{{\tilde \theta }^{\rm{T}}}\tilde \theta } \right\} + \\
 +tr\left\{ {{{\tilde \theta }^{\rm{T}}}\left( 0 \right)\omega {\omega ^{\rm{T}}}\tilde \theta \left( 0 \right) + {{\tilde \theta }^{\rm{T}}}\left( 0 \right)\tilde \theta \left( 0 \right)} \right\}.
\end{array}
\end{equation}

A maximum eigenvalue of $\omega\left( t \right)  {\omega ^{\rm{T}}}\left( t \right) $ over $\left[ {0{\rm{; }}\;\;{t_e}} \right)$ is introduced:
\begin{equation}\label{eqA6}
\delta  = {\rm{sup}}\mathop {{\rm{max}}}\limits_{\forall t < {t_e}} {\lambda _{{\rm{max}}}}\left( {\omega \left( t \right){\omega ^{\rm{T}}}\left( t \right)} \right).
\end{equation}

Taking into consideration \eqref{eqA6}, the equation \eqref{eqA5} for $t < {t_e}$ is rewritten as:
\begin{equation}\label{eqA7}
\begin{gathered}
\dot V \le  - \mu {\lambda _{{\rm{min}}}}\left( P \right){\left\| {{e_{ref}}} \right\|^2} - {\left\| {\tilde \theta } \right\|^{\rm{2}}} +\\
+\left( {\delta  + 1} \right){\left\| {\tilde \theta \left( 0 \right)} \right\|^2} \le  - {\eta _{\rm{1}}}V + {r_B}{\rm{,}}
\end{gathered}
\end{equation}
where ${\eta _{\rm{1}}} = {\rm{min}}\left\{ {{\textstyle{{\mu {\lambda _{{\rm{min}}}}\left( P \right)} \over {{\lambda _{{\rm{max}}}}\left( P \right)}}}{\rm{; 2}}} \right\}{\rm{; }}\;\;{r_B} = \left( {\delta  + 1} \right){\left\| {\tilde \theta \left( 0 \right)} \right\|^2}$.


\addtolength{\textheight}{-0cm}   


Having solved \eqref{eqA7}, it is obtained:
\begin{equation}\label{eqA8}
\forall t < {t_e}{\rm{: }}\;V \le {e^{ - {\eta _{\rm{1}}}t}}V\left( 0 \right) + {\textstyle{{{r_B}} \over {{\eta _{\rm{1}}}}}}.
\end{equation}

As ${\lambda _{\mathop{\rm m}\nolimits} }{\left\| \xi  \right\|^2} \le V$ and $V\left( 0 \right) \le {\lambda _M}{\left\| {\xi \left( 0 \right)} \right\|^2}$, then we obtain from \eqref{eqA8} that the estimate of $\xi\left( t \right) $ is bounded for $\forall t < {t_e}$:
\begin{equation}\label{eqA9}
\begin{gathered}
\left\| \xi\left( t \right)   \right\| \le \sqrt {{\textstyle{{{\lambda _M}} \over {{\lambda _m}}}}{e^{ - {\eta _{\rm{1}}}t}}{{\left\| {\xi \left( 0 \right)} \right\|}^2} + {\textstyle{{{r_B}} \over {{\lambda _m}{\eta _{\rm{1}}}}}}}  \le\\
\le \sqrt {{\textstyle{{{\lambda _M}} \over {{\lambda _m}}}}{{\left\| {\xi \left( 0 \right)} \right\|}^2} + {\textstyle{{{r_B}} \over {{\lambda _m}{\eta _{\rm{1}}}}}}} .
\end{gathered}
\end{equation}

Secondly, let $t \ge {t_e}$. Considering \eqref{eq21} and that   $\forall t \ge {t_e}$ $0 \!<\! {\Omega _{{\rm{LB}}}} \!\le\! \Omega\left( t \right)  \! \le \! {\Omega _{{\rm{UB}}}}$ holds, we obtain from \eqref{eq21} $\forall t \ge {t_e}$:
\begin{equation}\label{eqA10}
\begin{array}{c}
\dot V \le  - \mu e_{ref}^{\rm{T}}P{e_{ref}} + \\
+ tr\left\{ {{{\tilde \theta }^{\rm{T}}}\omega {\omega ^{\rm{T}}}\tilde \theta  - {{\tilde \theta }^{\rm{T}}}{\textstyle{{\left( {{\gamma _0}{\lambda _{{\rm{max}}}}\left( {\omega {\omega ^{\rm{T}}}} \right) + {\gamma _1}} \right){\Omega ^2}} \over {{\Omega ^{2}}}}}\tilde \theta } \right\} = \\
 =  - \mu e_{ref}^{\rm{T}}P{e_{ref}} + \\
 + tr\left\{ {{{\tilde \theta }^{\rm{T}}}\omega {\omega ^{\rm{T}}}\tilde \theta  - {{\tilde \theta }^{\rm{T}}}\left[ {{\gamma _0}{\lambda _{{\rm{max}}}}\left( {\omega {\omega ^{\rm{T}}}} \right) + {\gamma _1}} \right]\tilde \theta } \right\}.
\end{array}
\normalsize
\end{equation}

It holds for any $\omega\left( t \right) $ that:
\begin{equation}\label{eqA11}
\begin{gathered}
{\omega\left( t \right) {\omega ^{\rm{T}}\left( t \right)} - {\gamma _0}{\lambda _{{\rm{max}}}}\left( {\omega\left( t \right) {\omega ^{\rm{T}}}\left( t \right)} \right)I} \le - \kappa \le 0.
\end{gathered}
\end{equation}

So \eqref{eqA10} is written as:
\begin{equation}\label{eqA12}
\begin{gathered}
\dot V \le  - \mu e_{ref}^{\rm{T}}P{e_{ref}} - \left( {\kappa  + {\gamma _1}} \right)tr\left\{ {{{\tilde \theta }^{\rm{T}}}\tilde \theta } \right\} \le \\
 \le  - \mu {\lambda _{{\rm{min}}}}\left( P \right){\left\| {{e_{ref}}} \right\|^2} - \left( {\kappa  + {\gamma _1}} \right){\left\| {\tilde \theta } \right\|^{\rm{2}}} \le  - {\eta _{\rm{2}}}V{\rm{,}}
\end{gathered}
\end{equation}
where ${\eta _{\rm{2}}} = {\rm{min}}\left\{ {{\textstyle{{\mu {\lambda _{{\rm{min}}}}\left( P \right)} \over {{\lambda _{{\rm{max}}}}\left( P \right)}}}{\rm{;\;\;2}}\left( {\kappa  + {\gamma _1}} \right)} \right\}.$

The inequality \eqref{eqA12} is solved, and it is obtained for $t \ge {t_e}$:
\begin{equation}\label{eqA13}
V \le {e^{ - {\eta _2}t}}V\left( {{t_e}} \right).
\end{equation}

Considering ${\lambda _{\mathop{\rm m}\nolimits} }{\left\| \xi  \right\|^2} \le V$, $V\left( {{t_e}} \right) \le {\lambda _M}{\left\| {\xi \left( {{t_e}} \right)} \right\|^2}$ and \eqref{eqA9}, the estimate of $\xi\left( t \right) $ for  $t \ge {t_e}$ is obtained from \eqref{eqA13}:
\begin{equation}\label{eqA14}
\begin{gathered}
\left\| \xi\left( t \right)   \right\| \le \sqrt {{\textstyle{{{\lambda _M}} \over {{\lambda _m}}}}{e^{ - {\eta _2}t}}{{\left\| {\xi \left( {{t_e}} \right)} \right\|}^2}}  \le \\
\le \sqrt {{\textstyle{{{\lambda _M}} \over {{\lambda _m}}}}\left( {{\textstyle{{{\lambda _M}} \over {{\lambda _m}}}}{{\left\| {\xi \left( 0 \right)} \right\|}^2} + {\textstyle{{{r_B}} \over {{\lambda _m}{\eta _{\rm{1}}}}}}} \right)} .
\end{gathered}
\end{equation}

Hence, it is concluded from \eqref{eqA9} and \eqref{eqA14} that $\xi\left( t \right)   \in {L_\infty }$ and $\xi\left( t \right) $ converges to zero exponentially $\forall t \ge {t_e}$. The rate of such convergence is adjustable by $\gamma _0$, $\gamma _1$. Q.E.D.

\bibliography{ifacconf}             
                                                   







\end{document}